\documentclass[aps,pra,superscriptaddress,amsmath,amssymb,preprintnumbers,twocolumn,floats,floatsfix,showpacs]{revtex4}
\usepackage{amssymb}
\usepackage{graphicx}
\renewcommand{\eqref}[1]{Eq.(\ref{#1})}

\begin{document}

\title{Microscopic derivation of the Jaynes-Cummings model with cavity losses}

\author{M. Scala}
\email{matteo.scala@fisica.unipa.it}

\affiliation{MIUR and Dipartimento di Scienze Fisiche ed
Astronomiche dell'Universit\`{a} di Palermo, via Archirafi 36,
I-90123 Palermo, Italy}

\author{B. Militello}

\affiliation{MIUR and Dipartimento di Scienze Fisiche ed
Astronomiche dell'Universit\`{a} di Palermo, via Archirafi 36,
I-90123 Palermo, Italy}

\author{A. Messina}

\affiliation{MIUR and Dipartimento di Scienze Fisiche ed
Astronomiche dell'Universit\`{a} di Palermo, via Archirafi 36,
I-90123 Palermo, Italy}

\author{J. Piilo}

\affiliation{Department of Physics, University of Turku, FI-20014
Turku, Finland}

\author{S. Maniscalco}

\affiliation{Department of Physics, University of Turku, FI-20014
Turku, Finland}

\begin{abstract}
In this paper we provide a microscopic derivation of the master
equation for the Jaynes-Cummings model with cavity losses. We
single out both the differences with the phenomenological master
equation used in the literature and the approximations under which
the phenomenological model correctly describes the dynamics of the
atom-cavity system. Some examples wherein the phenomenological and
the microscopic master equations give rise to different
predictions are discussed in detail.
\end{abstract}


\maketitle

\section{Introduction}
The Jaynes-Cummings (JC) model is the fundamental model for the
quantum description of matter-light interaction~\cite{JC}. It
describes the dynamics of a two-level atom strongly interacting
with a single mode of the quantized radiation field in the
rotating-wave approximation (RWA). This model has been extensively
studied in the past three decades~\cite{shore_review, puribook}.
Purely quantum effects predicted by the model, such as Rabi
oscillations and collapses and revivals of the atomic inversion
operator, have been observed in the experiments both with
microcavities~\cite{haroche, harocheRMP, Walther} and with trapped
ion systems~\cite{wineland}.

In cavity quantum electrodynamics, the JC model describes the
strong-coupling regime of micromasers and of one-atom lasers. A
realistic description of these systems, however, must take into
account the photon losses due to imperfect reflectivity of the
cavity mirrors. Cavity losses have been described in the
literature by means of a phenomenological master equation of the
form
\begin{eqnarray}\label{oldME}
 \dot{\rho}=-i\left[H_{JC},\rho\right]+\gamma\left( a\rho
 a^\dag-\frac{1}{2}a^\dag a \rho -\frac{1}{2} \rho a^\dag
a\right),
\end{eqnarray}
where $\rho$ is the density matrix of the atom-cavity system,
$H_{JC}$ is the Jaynes-Cummings Hamiltonian, and $\gamma$
represents the rate of loss of photons from the cavity, and where
we have put $\hbar =1$. The second term on the r.h.s.~of
Eq.~(\ref{oldME}) has been derived microscopically in the
framework of a different physical problem, i.e.,~the one wherein
the system is given by the quantized cavity mode only, losing
excitations because of its interaction with the surrounding
electromagnetic field in the vacuum state~\cite{cohen_libro}.

In this paper we will show that, when the system consists not only
of the cavity mode, but also of the two-level atom interacting
with the cavity, the fully microscopic derivation gets more
complicated, and in general the master equation for the
atom-cavity system is not of the form given by Eq.~(\ref{oldME}).
We will show that only under certain conditions the
phenomenological master equation coincides with the microscopic
one. These conditions are typically met in the cavity QED
experiments, and this explains the extensive and successful use of
Eq.~(\ref{oldME}) for the description of the dynamics of the JC
model with losses. The knowledge of the limitations of the
phenomenological model, however, allows us on the one hand to
point out some misuse of the model, e.g., for the study of the
damping of highly excited quasi-classical states~\cite{geaben}. On
the other hand it paves the way to the correct description of
photon losses in the case of structured reservoirs, e.g. for
photonic bandgap cavities. Moreover, our derivation brings to
light the microscopic processes which occur in the open system
dynamics, and therefore how decoherence and dissipation come into
play.

The paper is structured as follows. In Sec.~II we give a short
review on the JC model and on the usual phenomenological
description of losses in this model. In Sec.~III we present the
microscopic derivation of the JC model with cavity losses, for the
general case of a $T$ temperature electromagnetic reservoir. In
Sec.~IV we describe the dynamics for the case of one initial
excitation in the atom-cavity system, we solve the master equation
and we compare it with the solution of the phenomenological model
existing in the literature. In Sec.~V we briefly discuss some
situations in which the use of the microscopic model could lead to
substantial differences compared to the phenomenological model,
suggesting experimental situations in which such differences could
be observed, and finally we present conclusions.

\section{The Jaynes-Cummings model and the phenomenological description of losses}

Since its introduction, in 1963~\cite{JC}, the JC model has been
one of the most used models for the description of
radiation-matter interaction in quantum
optics~\cite{shore_review}, in particular in cavity quantum
electrodynamics~\cite{haroche}, and in ion traps~\cite{wineland}.
In this section we briefly review some of its  features.

Let us consider a two-level atom and denote by $\left|g\right>$
and $\left|e\right>$ the ground and excited state, respectively.
The energy separation between the two states is given by $\hbar
\omega_0$, with $\omega_0$ the Bohr frequency. The resonant
interaction between the atom and a mode of the electromagnetic
field, in the RWA and in units of $\hbar$, is described by the
following Hamiltonian~\cite{JC}:
\begin{eqnarray}\label{JCM}
 H_{JC}=\frac{\,\omega_0}{2}\sigma_z+\omega_0\,a^\dag
 a+\Omega\left(a\sigma_++a^\dag\sigma_-\right),
\end{eqnarray}
where $a^\dag$ ($a$) is the creation (annihilation) operator of
the mode, $\sigma_-=\left|g\right>\left<e\right|$,
$\sigma_+=\left|e\right>\left<g\right|$, and $\sigma_z =
\left|e\rangle \langle e \right| - \left|g \rangle \langle g
\right|$.

It is straightforward to show that the total number of excitations
in the atom-cavity system, given by ${\cal N}=\left<a^\dag a
+\sigma_z+1/2\right>$, is a constant of motion. This allows to
 diagonalize easily the Hamiltonian $H_{JC}$. One finds the
following eigenstates and eigenvalues~\cite{shore_review}:
\begin{eqnarray}\label{eigenstates}
 &&\left|E_{N,\pm}\right>=\frac{1}{\sqrt{2}}\left(\left|N,g\right>\pm
 \left|N-1,e\right>\right),\nonumber\\
 \nonumber\\
 &&E_{N,\pm}=\left(N-\frac{1}{2}\right)\omega_0\pm\Omega\sqrt{N},
\end{eqnarray}
for $N \ge 1$, while the ground state and the corresponding energy
eigenvalue are
\begin{eqnarray}\label{groundeigenstate}
 \left|E_0\right>=\left|0,g\right>, \hspace{1cm}
 E_0=-\frac{\omega_0}{2},
\end{eqnarray}
respectively, where $\vert N, i \rangle = \vert N \rangle \vert i
\rangle$, with $i=e,g$, indicates the tensor product of the Fock
state $\vert N \rangle$ and the electronic states $\vert i
\rangle$.

From the eigenstates and the eigenvalues of $H_{JC}$ one can
calculate the evolution of the system given any initial
conditions. Well known examples of system dynamics are Rabi
oscillations of the atomic state population, and collapses and
revivals of the oscillations when the mode is initially in a
coherent state~\cite{shore_review}.

In cavity quantum electrodynamics the main source of dissipation
originates from the leakage of cavity photons due to imperfect
reflectivity of the cavity mirrors. A second source of dissipation
and decoherence, namely spontaneous emission of photons by the
atom, is mostly suppressed by the presence of the cavity, and
therefore its effect is usually neglected.

The dissipative dynamics of the quantized modes of the radiation
field inside the cavity {\it in absence of the atom}, i.e.,~when
the atom is not inside the cavity, can be derived microscopically
assuming that the cavity modes are coupled with the
electromagnetic field outside the cavity, which represents a
reservoir at temperature $T$~\cite{cohen_libro}. The typical
approach to the description of the losses in the JC model consists
in assuming that the presence of the atom inside the cavity does
not modify strongly the mechanism of cavity losses, which
therefore can be modelled by the above mentioned master equation.
It is worth stressing further, however, that this approach is
purely phenomenological, since it does not take into account, in
the microscopic derivation, the presence of the atom inside the
cavity. The phenomenological master equation has the form
\begin{eqnarray}\label{phenME}
 \dot{\rho}&=&-i\left[H_{JC},\rho\right] \nonumber\\
\nonumber\\&+& \gamma \left[ n(\omega_0)+1 \right]  \left[ a\rho
 a^\dag - \frac{1}{2}\left(a^\dag a \rho+\rho a^\dag
a \right)\right]\nonumber\\
\nonumber\\
&+& \gamma n(\omega_0)  \left[ a^\dag\rho
 a-\frac{1}{2}\left(a a^\dag \rho+\rho a
a^\dag\right)\right],
\end{eqnarray}
with $n(\omega_0)$ the average number of quanta of the reservoir
in the mode of frequency $\omega_0$, and $\gamma$ the rate of loss
of cavity photons. We note that for a zero-$T$ reservoir the
master equation above reduces to the one given by
Eq.~(\ref{oldME}). Equations (\ref{oldME}) and~(\ref{phenME}) have
been assumed to be valid in most of the earlier studies dealing
with the JC model with losses, for example in
\cite{agar1,agar2,agar3,barnett,briegel,puribook,harocheRMP}.

The use of the master equations given by Eqs.~(\ref{oldME})
and~(\ref{phenME}) has also been motivated by the fact that in
many cavity QED experiments the atoms fly through the cavity and
actually remain inside the cavity only for a short time. This
could induce one to think that the effect of their presence inside
the cavity might be negligible. We believe, however, that a
comparison with a microscopic master equation describing the
coupling of the entire atom-cavity system with a reservoir of
electromagnetic modes at $T$ temperature is highly desirable and
it may both provide a justification of the validity of the
phenomenological model under the typical experimental conditions
and put into evidence the physical contexts where the use of such
a model may be unjustified. This is the main motivation of the
results we will describe in the following sections.

We begin, in the next section, by presenting the general
formalism, reviewed,  e.g., in Ref.~\cite{petruccionebook}, to
derive a master equation for the open quantum system of interest
starting from the microscopic Hamiltonian of the total closed
system (system+environment). We will then compare the master
equation we obtain with the one given by Eq.~(\ref{phenME}).

\section{Microscopic derivation of the quantum master equation}

\subsection{The general formalism}

We assume that the open quantum system of interest, e.g., the
atom-cavity system, is  part of a larger system whose dynamics is
unitary and governed by the Hamiltonian $H$. The external
environment is that part of the total closed system other than the
system of interest. The Hamiltonian of the total closed system is
given by
\begin{eqnarray}
\label{general H}
 H=H_S+H_E+H_{\rm int},
\end{eqnarray}
where $H_S$ and $H_E$, are the system and environment
Hamiltonians, respectively, and $H_{\rm int}$ is the
system-environment interaction Hamiltonian which is taken to be of
the form
\begin{eqnarray}\label{interaction}
 H_{\rm int}=A\otimes E,
\end{eqnarray}
with $A=A^\dag$ and $E=E^\dag$  Hermitian operators acting on the
system and on the environment Hilbert spaces, respectively.

We expand the interaction Hamiltonian $H_{\rm int}$ by means of
the relations
\begin{eqnarray}
\label{Aomega}
 A&=&\sum_{\omega} A(\omega),\nonumber \\
 A(\omega)&=&\sum_{\epsilon'-\epsilon=\omega}\Pi(\epsilon)A\Pi(\epsilon'),
\end{eqnarray}
where $\Pi(\epsilon)$ is the projector onto the eigenspace
corresponding to the eigenvalue $\epsilon$ of the operator $H_S$
and the sum is taken over all the Bohr frequencies relative to
$H_S$.

Following the standard procedure, i.e.,~writing down the
Liouville-von Neumann equation for the total density operator in
the interaction picture with respect to $H_S+H_E$, performing the
Born-Markov and the rotating wave approximations, tracing out the
environmental degrees of freedom and then going back to the
Schr\"odinger picture, one obtains the following master equation
for the reduced density operator $\rho$ of the
system~\cite{petruccionebook}:
\begin{eqnarray}\label{meq_general}
 &&\dot{\rho}(t)=-i[H_S,\rho(t)]\nonumber\\
 &&+\sum_{\omega>0}\gamma(\omega)\left[A(\omega)\rho(t)A^\dag(\omega)-\frac{1}{2}\left\{A^\dag(\omega)A(\omega),\rho(t)\right\}\right] \nonumber \\
 &&+\!\!\!\!\sum_{\omega>0}\!\!\!\gamma(-\omega)\!\left[A^\dag(\omega)\rho(t)A(\omega)\!-\!\frac{1}{2}\left\{A(\omega)A^\dag(\omega),\rho(t)\right\}\right]\!,
\end{eqnarray}
where  the relation $A(-\omega)=A^\dag(\omega)$ has been used and
where we have neglected the renormalization
term~\cite{petruccionebook}. The coefficients $\gamma(\omega)$ are
given by the Fourier transform of the correlation functions of the
environment:
\begin{eqnarray}
 \gamma(\omega)=\int_{-\infty}^{+\infty}d\tau\,
 \mathrm{e}^{i\omega\tau}\left<E^\dag(\tau)E(0)\right>,\label{gammaomega}
\end{eqnarray}
where the environment operators are in the interaction picture.

We stress once more that the formalism presented above is valid as
long as we can perform three approximations: weak coupling or Born
approximation, Markovian approximation and rotating wave
approximation. It is worth recalling that the Markovian
approximation can be seen as a coarse-graining in time, and
therefore holds as long as the correlation time of the reservoir
$\tau$ is much smaller than the characteristic time scale of the
system dynamics $t$
\begin{eqnarray}\label{markov}
 t\gg \tau.
\end{eqnarray}
The RWA, instead, is valid as long as the relaxation time of the
system is much longer than the typical timescale of the free
evolution of the quantum system, i.e.,~when the maximum of the
rates $\gamma(\omega)$ is much smaller than the minimum difference
between the Bohr frequencies relative to $H_S$
\cite{petruccionebook}:
\begin{eqnarray}\label{RWA}
\gamma_{\rm max}\ll\Delta\omega_{\rm min}.
\end{eqnarray}
In the rest of this section we will apply this general formalism
to the description of cavity losses in the Jaynes-Cummings model.

\subsection{Application of the general formalism to the Jaynes-Cummings model}
We model the environment as a collection of quantum harmonic
oscillators in thermal equilibrium at $T$ temperature and we
assume that the interaction Hamiltonian is linear in both the
electromagnetic field of the cavity mode and the position
operators of the harmonic oscillators, i.e.,
\begin{eqnarray}\label{modello}
 H_S&=&H_{JC},\;\;\;H_E=\sum_k\omega_k b^\dag b, \nonumber\\
            \nonumber\\
            H_{\rm int}&=&\left(a+a^\dag\right)\sum_kg_k\left(b_k+b_k^\dag\right),
\end{eqnarray}
with $\omega_k$ the frequencies of the environment oscillators,
$b^{\dag}_k$ ($b_k$) the creation (annihilation) operator of
quanta in the $k$-th environmental mode, and $g_k$ the coupling
constants.

For this system, the operators $A(\omega)$, defined in
Eq.~(\ref{Aomega}), are given by
\begin{eqnarray}\label{aomegabis}
 &A&\!\!\!\left(\!E_{N'\!,\,l}\!-\!E_{N,\,m}\right)=\left|E_{N,\,m}\right>\left<E_{N,\,m}\right|\!\left(a+a^\dag\right)\!
                        \left|E_{N'\!,\,l}\right>\left<E_{N'\!,\,l}\right| \nonumber \\
                        &=&\frac{1}{2}\delta_{N,N'-1}\!\left(\!\sqrt{N+1}+lm\sqrt{N}\right)\left|E_{N,\,m}\right>\left<E_{N+1,\,l}\right|,
\end{eqnarray}
for $N\ge1$ and
\begin{eqnarray}\label{aomegabisN1}
A\left(E_{1,\pm}-E_0\right)=\frac{1}{\sqrt{2}}\left|E_0\right>\left<E_{1,\pm}\right|,
\end{eqnarray}
for $N=1$. In Eq.~(\ref{aomegabis}) we indicate the states
$\left|E_{N,\pm}\right>$ by $\left|E_{N,\pm 1}\right>$ and the
energy eigenvalues $E_{N,\pm}$ by $E_{N,\pm 1}$. Accordingly $l$
and $m$ take the values $\pm1$.

Having in mind Eq.~(\ref{meq_general}) we can write the Markovian
RWA master equation for the JC model interacting with a thermal
bath at $T$ temperature as follows
\begin{widetext}
\begin{eqnarray}\label{equazioneT}
 \dot{\rho}&=&-i\left[H_{JC},\rho\right]
              +\sum_{l=\pm1}\frac{\gamma\left(E_{1,\,l}-E_0\right)}{2}\left(\left|E_{0}\right>\left<E_{1,\,l}\right|\rho
              \left|E_{1,\,l}\right>\left<E_0\right|-\frac{1}{2}\left\{\left|E_{1,\,l}\right>\left<E_{1,\,l}\right|,\rho\right\}\right)\nonumber\\
              &+&\sum_{l,m=\pm 1}\sum_{N=1}^{\infty}\frac{\gamma\left(E_{N+1,\,l}-E_{N,\,m}\right)}{4}\left(\sqrt{N+1}+lm\sqrt{N}\right)^2
              \Big(\left|E_{N,\,m}\right>\left<E_{N+1,\,l}\right|\rho\left|E_{N+1,\,l}\right>\left<E_{N,\,m}\right|\nonumber\\
              &-&\left.
              \frac{1}{2}\left\{\left|E_{N+1,\,l}\right>\left<E_{N+1,\,l}\right|,\rho\right\}\right)\nonumber\\
              &+&\sum_{l=\pm1}\frac{\gamma\left(E_0-E_{1,\,l}\right)}{2}\left(\left|E_{1,\,l}\right>\left<E_{0}\right|\rho
              \left|E_0\right>\left<E_{1,\,l}\right|-\frac{1}{2}\left\{\left|E_0\right>\left<E_0\right|,\rho\right\}\right)\nonumber \\
               &+&\sum_{l,m=\pm 1}\sum_{N=1}^{\infty}\frac{\gamma\left(E_{N,\,m}-E_{N+1,\,l}\right)}{4}\left(\sqrt{N+1}+lm\sqrt{N}\right)^2
              \Big( \left|E_{N+1,\,l}\right>\left<E_{N,\,m}\right|\rho\left|E_{N,\,m}\right>\left<E_{N+1,\,l}\right|\nonumber\\
              &-&\left.
              \frac{1}{2}\left\{\left|E_{N,\,m}\right>\left<E_{N,\,m}\right|,\rho\right\}\right).
\end{eqnarray}
\end{widetext}
The Kubo-Martin-Schwinger condition \cite{kms}
\begin{eqnarray}\label{KMS}
 \gamma(-\omega)=\mathrm{exp}\left(-\frac{\omega}{k_BT}\right)\gamma(\omega),
\end{eqnarray}
ensures that the stationary state reached at time $t=+\infty$ is
the thermal state~\cite{petruccionebook}
\begin{eqnarray}\label{thermal new}
    \rho_{\rm
    th}=\frac{\mathrm{exp}\left(-\frac{H_{JC}}{k_BT}\right)}{Tr\left\{\mathrm{exp}\left(-\frac{H_{JC}}{k_BT}\right)\right\}},
\end{eqnarray}
as expected from statistical mechanical considerations and as can
be easily justified by the detailed balance principle and by means
of Eq.~(\ref{KMS}).

It is worth underlining a first difference between our microscopic
master equation, given by Eq.~(\ref{equazioneT}), and the
phenomenological one given by Eq.~(\ref{phenME}). While the
thermal stationary state predicted by the phenomenological model
is given by
\begin{eqnarray}\label{thermal old}
 \rho_{\rm th}^{\rm ph}=\frac{\mathrm{exp}\left(-\frac{\frac{\,\omega_0}{2}\sigma_z+\omega_0\,a^\dag
 a}{k_BT}\right)}{Tr\left\{\mathrm{exp}\left(-\frac{\frac{\,\omega_0}{2}\sigma_z+\omega_0\,a^\dag
 a}{k_BT}\right)\right\}},
\end{eqnarray}
our microscopic approach predicts that the stationary state is the
one given by Eq.~(\ref{thermal new}), which differs from
Eq.~(\ref{thermal old}) for the presence of the interaction energy
term in $H_{JC}$ [See Eq.~(\ref{JCM})].

We conclude this section noting the limit of validity of our
microscopic master equation. From Eqs.~(\ref{eigenstates}) and
(\ref{RWA}) we deduce that the RWA we have performed is valid as
long as the smallest difference between Bohr frequencies relative
to $H_{JC}$ is much larger than the highest decay rate of the
system, i.e.
\begin{eqnarray}\label{RWA_JC}
 2\Omega\gg\gamma_{\rm max},
\end{eqnarray}
since the typical evolution timescale of the system is given by
the inverse of the Rabi frequency $2\Omega$.

\section{Comparison between the two master equations}

\subsection{The phenomenological master equation in the dressed-state approximation and its relation with the experiments}

In the first part of this section we examine an important feature
of the phenomenological master equation given by Eq.~(\ref{oldME})
which explains its success in describing accurately most of the
cavity QED experiments.

In~\cite{agar1,agar2,barnett} it has been shown that, in a regime
very close to the one in which our microscopic master equation is
valid, i.e. for $\gamma\ll\Omega$, one can approximate the
phenomenological master equation by means of the so-called {\em
dressed-state approximation}.  In a later paper~\cite{geaben} the
validity of this approximation was carefully analyzed and it was
discovered that it actually requires the stronger condition
$\gamma\ll \Omega / (2 N^{3/2})$ in order to be valid. The
dressed-state approximation amounts at neglecting, in the
interaction picture with respect to the Hamiltonian $H_{JC}$, all
the time-dependent terms oscillating at frequencies which are
multiple of $\Omega$, under the hypothesis that this frequency is
much larger than the decay rate $\gamma$.

In the Schr\"odinger picture, the set of coupled differential
equations for the matrix elements, in the dressed-state
approximation,
$\left<E_{N,+}\right|\dot{\rho}\left|E_{N,+}\right>$,
$\left<E_{N,-}\right|\dot{\rho}\left|E_{N,-}\right>$ and
$\left<E_{N,\pm}\right|\dot{\rho}\left|E_{N,\mp}\right>$ coincide
with the corresponding set of equations obtained from our master
equation~(\ref{equazioneT}) at zero temperature, {\em when the
spectrum of the environment is flat, i.e. in the case of white
noise}. In other words, our microscopic approach justifies the
validity of the dressed-state approximation in terms of a
microscopic system-reservoir interaction model, and explains the
success of the phenomenological master equation in fitting the
experimental data in the strong coupling regime, i.e.,~when the
relaxation time is much longer than the frequency of the Rabi
oscillations~\cite{harocheRMP}. It is worth stressing, however,
that if the spectrum of the environment is not flat the
predictions of the two master equations differ also in the limit
of weak damping since, in this case, the differential equations
for the relevant matrix elements obtained from our microscopic
master equation do not coincide with the differential equations
obtained from the phenomenological master equation after
performing the dressed-state approximation~\cite{agar1,agar2}.

In the next subsection we present a comparison between the
predictions of the phenomenological and microscopic master
equations at zero temperature when the system has one initial
excitation. We will concentrate on a finite (three) dimensional
subspace using the phenomenological model without the
dressed-state approximation, because the restriction of
Eq.~(\ref{oldME}) to such a subspace gives rise to an exactly
solvable dynamical case.

Since our microscopic model gives the same predictions of the
dressed-state approximation for the phenomenological model, the
examples we are going to consider in the next subsection will give
us also insight in the limits of validity of the dressed-state
approximation itself.

As we will see in the following, the discrepancy between the
phenomenological master equation (without dressed-state
approximation) and the microscopic master equation is indeed
appreciable already at the first order in $\gamma/\Omega$.
Consequently, for the value of the parameters considered in our
examples, the applicability of the dressed state approximation to
the phenomenological model appears to be questionable.

\subsection{Dynamics at $T=0$ with one initial excitation}

\subsubsection{Decay of the Rabi oscillations with the atom initially
excited}

We assume the initial state of the system is $\left|0,e\right>$.
In absence of cavity losses one would observe a continuous
exchange of energy between the atom and the cavity mode, namely
the Rabi oscillations. The interaction between the cavity and the
environment causes the loss of energy from the atom-cavity system
to the external environment. Since in the system there is only one
initial excitation and the environment is at zero temperature, the
number of excitations cannot increase in time and
Eq.~(\ref{equazioneT}) reduces to the following simplified master
equation:
\begin{eqnarray}\label{equation1excitation}
 \dot{\rho}&=&-i\left[H_{JC},\rho\right]\nonumber\\
\nonumber\\&+&\gamma\left(\omega_0+\Omega\right)
  \left(\frac{1}{2}\left|E_0\right>\left<E_{1,+}\right|\rho\left|E_{1,+}\right>\left<E_0\right|\right.\nonumber\\
\nonumber\\&-&\left.\frac{1}{4}\left\{\left|E_{1,+}\right>\left<E_{1,+}\right|,\rho\right\}\right)\nonumber\\
\nonumber\\
 &+&\gamma\left(\omega_0-\Omega\right)
  \left(\frac{1}{2}\left|E_0\right>\left<E_{1,-}\right|\rho\left|E_{1,-}\right>\left<E_0\right|\right.\nonumber\\
\nonumber\\
&-&\left.\frac{1}{4}\left\{\left|E_{1,-}\right>\left<E_{1,-}\right|,\rho\right\}\right),
\end{eqnarray}
obtained from Eq.~(\ref{equazioneT}) neglecting all the terms which
do not contribute to the evolution of the system.

In the Appendix we provide the solution of both Eq.~(\ref{oldME})
and Eq.~(\ref{equation1excitation})  based on the method of the
damping basis~\cite{briegel}.
\begin{figure}\label{fig1}
\begin{center}
     \includegraphics[ width=0.46\textwidth, height=0.28\textwidth ]{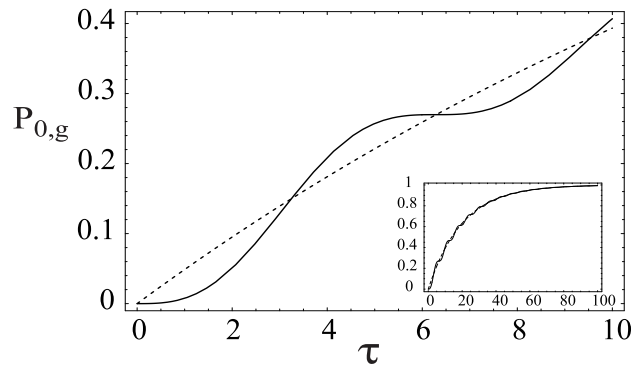}
    \caption{Populations $P_{0,g}(t)$ and $P_{0,g}^{\rm ph}(t)$
    vs $\tau=2\Omega t$, when the system starts from the state
$\left|0,e\right>$, with $\gamma/2\Omega=0.1$.
    The solid line refers to the predictions given by the phenomenological master equation,
    while the dashed line refers to the predictions given by the microscopic one.
    In the inset it is shown the long time behavior of the same quantities,
    for $0 \le \tau \le 100$.}
 \end{center}
\end{figure}
In Fig.~1 we plot the time evolution of the population of the
state $\left|0,g\right>$ predicted by both the phenomenological
and the microscopic master equations, with
$\gamma\left(\omega_0-\Omega\right)=\gamma\left(\omega_0+\Omega\right)=\gamma=\Omega/5$.
This condition legitimates the use of our RWA master equation,
while it does not assure the validity of the dressed-state
approximation on the phenomenological model, in accordance
to~\cite{geaben}. The populations are given by
\begin{equation}\label{0g0gRabinewequalcoeff}
P_{0,g}(t)= \left<0,g\right|\rho(t)\left|0,g\right>=1-\mathrm{e}^{-\frac{\gamma}{2}t},\\
\end{equation}
for the predictions of Eq.~(\ref{equation1excitation})
(microscopic model), as one can see from Eq.~(\ref{0g0gRabinew})
in the Appendix putting $\gamma_a=\gamma_b=\gamma$, and by
\begin{eqnarray}
&&P_{0,g}^{\rm ph}(t)=\left<0,g\right|\rho
(t)\left|0,g\right>=1-\frac{16\Omega^2}{16\Omega^2-\gamma^2}\mathrm{e}^{-\frac{\gamma}{2}t}\nonumber \\
  &&+\frac{\gamma^2+\gamma\sqrt{\gamma^2-16\Omega^2}}{2\left(16\Omega^2-\gamma^2\right)}\mathrm{e}^{\frac{-\gamma+\sqrt{\gamma^2-16\Omega^2}}{2}t}\nonumber\\
  &&+\frac{\gamma^2-\gamma\sqrt{\gamma^2-16\Omega^2}}{2\left(16\Omega^2-\gamma^2\right)}\mathrm{e}^{\frac{-\gamma-\sqrt{\gamma^2-16\Omega^2}}{2}t},\label{0g0gRabiold}
\end{eqnarray}
for the predictions of Eq.~(\ref{oldME}) (phenomenological master
equation).

In both cases, the ground state population of the atom-cavity
system increases in time with the same exponential rate, due to
the cavity losses. There is, however, an important difference in
the behavior predicted by the two equations. Indeed, while our
microscopic master equation predicts a purely exponential
increase, the phenomenological master equation predicts the
presence of oscillations at the Rabi frequency superimposed to the
exponential increase. As anticipated, the amplitude of these
oscillations is of the order of $\gamma/\Omega$. This difference
in the ground state population dynamics reflects different
physical mechanisms in the dissipation process. According to the
phenomenological master equation, given by Eq.~(\ref{oldME}), only
the cavity can directly lose excitations, as one can see from the
form of the dissipator. The rate of loss is indeed proportional to
the population of the state $\left|1,g\right>$ only, while the
state $\left|0,e\right>$ does not directly decay. In this sense
the oscillating behavior of the population of the ground state is
a signature of the Rabi oscillations which induce the system decay
via the coupling of the state $\left|0,e\right>$ to the state
$\left|1,g\right>$.

The microscopic master equation given by Eq.~(\ref{equazioneT}),
with
$\gamma\left(\omega_0-\Omega\right)=\gamma\left(\omega_0+\Omega\right)$,
predicts instead that both the states $\left|1,g\right>$ and
$\left|0,e\right>$ decay with the same rate, since they are both
superpositions of the states $\left|E_{1,+}\right>$ and
$\left|E_{1,-}\right>$. For this reason there is no oscillating
behavior in the time evolution of the population of
$\left|0,g\right>$.

The difference between the predictions of the phenomenological
master equation and those of the microscopic master equation can
be revealed by performing joint measurements on both the atom and
the cavity, in order to get the population of the state
$\left|0,g\right>$ of the composite system.

We can observe a difference in the system dynamics also measuring
the population of the atomic ground state, given by
$P_{g}(t)=\left<0,g\right|\rho(t)\left|0,g\right>+\left<1,g\right|\rho(t)\left|1,g\right>$.
This type of measurement can be performed with standard
techniques~\cite{harocheRMP}. With the help of
Eq.~(\ref{equation1excitation}) one finds [See
Eqs.~(\ref{0g0gRabinew}) and Eq.~(\ref{1g1gRabinew}) in the
Appendix]
\begin{eqnarray}\label{ggRabinew}
 P_g\left( t\right)=1-\frac{1}{2}\mathrm{e}^{-\frac{\gamma}{2}t}
 -\frac{1}{4}\left[\mathrm{e}^{\left(2i\Omega-\frac{\gamma}{2}\right)t}-\mathrm{e}^{\left(-2i\Omega-\frac{\gamma}{2}\right)t}\right],
\end{eqnarray}
while using Eq.~(\ref{oldME}) one finds, from Eq.
~(\ref{0g0gRabiold}) and Eq.~(\ref{1g1gRabiold}) in the Appendix,
\begin{eqnarray}\label{ggRabiold}
  &&P_g^{\rm ph}(t)=1-\frac{8\Omega^2}{16\Omega^2-\gamma^2}\mathrm{e}^{-\frac{\gamma}{2}t}\nonumber\\
\nonumber\\
  &&+\frac{2\gamma^2+2\gamma\sqrt{\gamma^2-16\Omega^2}-16\Omega^2}{4\left(16\Omega^2-\gamma^2\right)}
  \mathrm{e}^{\frac{-\gamma+\sqrt{\gamma^2-16\Omega^2}}{2}t}\nonumber\\
\nonumber\\
  &&+\frac{2\gamma^2-2\gamma\sqrt{\gamma^2-16\Omega^2}-16\Omega^2}{4\left(16\Omega^2-\gamma^2\right)}
  \mathrm{e}^{\frac{-\gamma-\sqrt{\gamma^2-16\Omega^2}}{2}t}.
\end{eqnarray}

In this case, however, the behavior predicted by the
phenomenological master equation is very similar to the one
predicted by the microscopic one, as one can see from Fig.~2.
\begin{figure}\label{fig2}
\begin{center}
     \includegraphics[ width=0.46\textwidth, height=0.28\textwidth ]{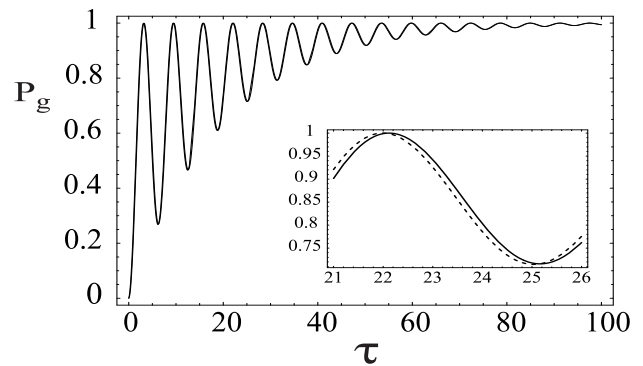}
    \caption{Populations $P_{g}(t)$ and $P_g^{\rm ph}(t)$ vs
    $\tau=2\Omega t$ for the initial state $\left|0,e\right>$, with $\gamma/2\Omega=0.1$.
    The solid line refers to the predictions given by the phenomenological master equation,
    while the dashed line refers to the predictions given by the microscopic one.
    In the large panel, where one can see the long time behavior, the two lines are indistinguishable,
    while in the inset we have emphasized the dynamics for $21 \le \tau \le 26$,
    to put into evidence the presence of the frequency shift.}
 \end{center}
\end{figure}
The only difference, indeed, is the presence of a frequency shift
in the Rabi oscillations, as predicted by the phenomenological
model. However, since such a shift is of the order of
$\left(\gamma/\Omega\right)^2$~\cite{agar3}, the differences in
the predicted behavior may be difficult to detect.

\subsubsection{Decay of a Bell state: exponential vs. oscillatory
behavior.}

We now consider the case in which the atom-cavity system is
initially prepared in a Bell state, such as the state
$\left|E_{1+}\right>$, as given by Eq.~(\ref{eigenstates}).
Contrarily to the case considered in the previous subsection, this
time the measurement of the atomic ground state population
$P_g(t)$ would allow to bring to light the differences in the
predictions of the phenomenological and microscopic master
equations.
\begin{figure}\label{fig3}
\begin{center}
     \includegraphics[ width=0.46\textwidth, height=0.28\textwidth ]{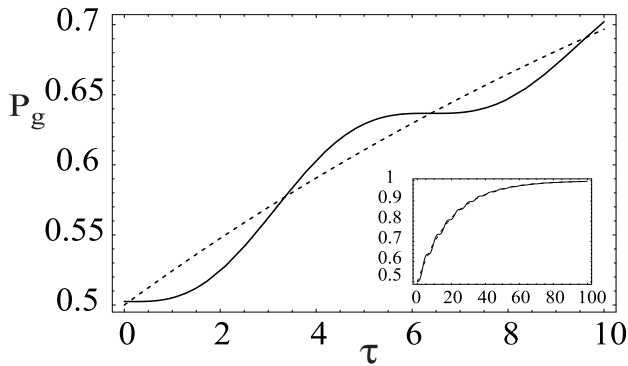}
    \caption{Populations $P_{g}(t)$ and $P_g^{\rm ph}(t)$ vs
    $\tau=2\Omega t$, when the system starts from the state $\left|E_{1,+}\right>$, with $\gamma/2\Omega=0.1$.
    The solid line refers to the predictions given by the phenomenological master equation,
    while the dashed line refers to the predictions given by the microscopic one.
    In the inset one can see the long time behavior of the same quantities for $0 \le \tau \le 100$.}
 \end{center}
\end{figure}
For the case of the initial state $\left|E_{1+}\right>$ the ground
state population is given by
\begin{equation}\label{bellpop}
 P_g\left( t  \right)=1-\frac{1}{2}\mathrm{e}^{-\frac{\gamma}{2}t},
\end{equation}
for the microscopic model [See Eq.~(\ref{equazioneT})] and
\begin{eqnarray}\label{ggBellold}
&&P_g^{\rm ph}(t)=1-\frac{8\Omega^2}{16\Omega^2-\gamma^2}\mathrm{e}^{-\frac{\gamma}{2}t}\nonumber \\
  &&+\frac{\gamma^2+\gamma\sqrt{\gamma^2-16\Omega^2}}{4\left(16\Omega^2-\gamma^2\right)}\mathrm{e}^{\frac{-\gamma+\sqrt{\gamma^2-16\Omega^2}}{2}t} \nonumber \\
  &&+\frac{\gamma^2-\gamma\sqrt{\gamma^2-16\Omega^2}}{4\left(16\Omega^2-\gamma^2\right)}\mathrm{e}^{\frac{-\gamma-\sqrt{\gamma^2-16\Omega^2}}{2}t},
\end{eqnarray}
for the phenomenological model [See Eq.~(\ref{oldME})]. Details on
the derivation of the above formulas are given in the Appendix.

As we see from Fig.~3, the atomic ground state population, as
predicted by the microscopic master equation given by
Eq.~(\ref{equazioneT}), exhibits a purely exponential behavior
while the phenomenological master equation, given by
Eq.~(\ref{oldME}), predicts the presence of oscillations at the
Rabi frequency with amplitude of the order of $\gamma/\Omega$. In
other words, in the microscopic approach, the atom-cavity system
as a whole is subjected to cavity losses and therefore the atom
dressed by the cavity mode can directly decay, although it is only
the cavity which is directly coupled to the environment.

We stress that, although the generation of a Bell state may be
more complicated than the generation of a Fock state, the set of
measurements necessary to distinguish between the predictions of
the two master equations when the initial state is
$\left|E_{1,+}\right>$ is simpler than the set of joint
measurements needed in the case of the initial state
$\left|0,e\right>$. In fact in the case of the Bell state we just
need a way to experimentally distinguish between the atomic
excited and ground state, for example by means of a static
electric field which ionizes the excited state only, as shown
in~\cite{harocheRMP}.

\section{Discussion and conclusive remarks}
In the previous section we have seen that our derivation of the
dissipative dynamics of the JC model with losses, taking into
account the presence of the atom inside the cavity, allows to give
a microscopic description of the interaction between the
atom-cavity open quantum system and the external electromagnetic
field.

It appears evident from our results that, although the source of
the losses is the escape of photons through the cavity mirrors, it
is the atom-cavity system as a whole, i.e. the atom dressed by the
cavity mode, which comes into play in the dissipative dynamics.
Stated another way, while the jump operators appearing in the
phenomenological master equations, given by Eq.~(\ref{oldME}), are
the annihilation and creation operators of photons in the cavity
($a$, and $a^{\dag}$), those appearing in the microscopic master
equation [See, e.g., Eq.~(\ref{equation1excitation})] describe
transitions between the dressed states of the atom-cavity system.

A noteworthy point emerging from our analysis is that the
deviations of the exactly solved phenomenological model from our
microscopic model (coinciding with the one obtained from the
phenomenological model with the dressed-state approximation) are
of the first order in $\gamma/\Omega$.  For this reason it should
be possible to check experimentally the predictions of the
microscopic model with the set up currently used in cavity QED
experiments.

It is worth stressing that in Eq.~(\ref{oldME}) only one cavity
loss rate $\gamma\equiv \gamma(\omega_0)$, with $\gamma(\omega)$
given by Eq.~(\ref{gammaomega}), appears. On the contrary, in
Eq.~(\ref{equation1excitation}), the loss rates corresponding to
jump operators connecting the ground state with the states $\vert
E_{1,+} \rangle$ and $\vert E_{1,-}\rangle$ are different, and
more precisely they are given by $\gamma(\omega_0+\Omega)$ and
$\gamma(\omega_0-\Omega)$, respectively. This clearly indicates
that, when the spectrum of the reservoir is not flat, the
phenomenological master equation does not provide a description
which can be justified in terms of a microscopic system-reservoir
interaction model as the one we have considered.

In the light of the considerations made above we expect that,
already in the weak coupling regime, differences between our
approach and the phenomenological one should become evident in
those physical contexts where the reservoir is structured, e.g. in
photonic band gap materials and hybrid solid-state cavity QED
systems~\cite{photbg1,photbg2,photbg3}.

\section*{Acknowledgements}
S.M. and J.P. acknowledge financial support from the Academy of
Finland (projects 206108, 108699), and the Magnus Ehrnrooth
Foundation.

M.S. thanks the Quantum Optics Group of the University of Turku
for the kind hospitality during the period May-August 2006.


\appendix
\section*{Appendix}
In this appendix we recall the method of the damping basis
introduced in Ref.~\cite{briegel} to solve master equations, and
we apply it to Eq.~(\ref{equation1excitation}).

Given a master equation of the form
\begin{equation}\label{lindbl}
 \dot{\rho}=\mathcal{L}\rho,
\end{equation}
where $\mathcal{L}$ is a time-independent linear superoperator
acting on $\rho$, one considers the following eigenvalue problem
\begin{equation}\label{righteigenprobl}
 \mathcal{L}\rho_\lambda=\lambda\rho_\lambda,
\end{equation}
where $\rho_\lambda$ is a right eigenoperator of the superoperator
$\mathcal{L}$ with eigenvalue $\lambda$. When the set of right
eigenoperators $\left\{\rho_\lambda\right\}$ is a basis for the
space of linear operators acting on the Hilbert space of the
system, as in the case analyzed in this paper, any density
operator can be expanded with respect to this set.

It is easy to show that if the system starts from the initial
condition
\begin{equation}\label{rhoinitial}
 \rho(0)=\sum_\lambda c_\lambda\rho_\lambda,
\end{equation}
then the time evolution of the system is then given by
\begin{equation}\label{rhofinal}
 \rho(t)=\sum_\lambda c_\lambda \mathrm{e}^{\lambda t}\rho_\lambda
\end{equation}

The coefficients $\left\{c_\lambda\right\}$ of the decomposition
in Eq.~(\ref{rhoinitial}) are given by
\begin{equation}\label{trace}
 c_\lambda=\mathrm{Tr}\left\{\check{\rho}_\lambda\,\rho(0)\!\right\},
\end{equation}
where $\check{\rho}_\lambda$ is the solution of the left
eigenvalue problem
\begin{equation}\label{lefteigenprobl}
 \check{\rho}_\lambda\mathcal{L}=\lambda\check{\rho}_\lambda,
\end{equation}
where the set of left eigenvalues $\left\{\lambda\right\}$ is the
same as in the right eigenvalue problem~\cite{briegel}.

When the dimension of the Hilbert space one considers is finite
and equal to $n$, the superoperator can be represented by a
non-hermitian $n^2\times n^2$ matrix.  The right eigenoperators,
if they exist, are represented by $n^2$-component column vectors,
while the left eigenoperators are represented by $n^2$-component
row vectors.

Applying this method to Eq.~(\ref{equation1excitation}), we have
to look for the nine right eigenoperators of its superoperator.
For simplicity we call
$\gamma_a=\gamma\left(\omega_0-\Omega\right)$ and
$\gamma_b=\gamma\left(\omega_0+\Omega\right)$.

Six of the nine right eigenoperators are given by the coherences
$\left|E_0\right>\left<E_{1,-}\right|$,
$\left|E_0\right>\left<E_{1,+}\right|$,
$\left|E_{1,-}\right>\left<E_{1,+}\right|$ and their hermitian
conjugates, with eigenvalues respectively equal to
$i\left(\omega_0-\Omega\right)-\gamma_a/2$,
$i\left(\omega_0+\Omega\right)-\gamma_b/2$,
$i(2\Omega)-\left(\gamma_a+\gamma_b\right)/4$ and their complex
conjugates. The other three right eigenoperators are given by
$\left|E_0\right>\left<E_0\right|$,
$\left(\left|E_{1,-}\right>\left<E_{1,-}\right|-\left|E_0\right>\left<E_0\right|\right)$,
$\left(\left|E_{1,+}\right>\left<E_{1,+}\right|-\left|E_0\right>\left<E_0\right|\right)$
with eigenvalues respectively equal to $0$, $-\gamma_a/2$ and
$-\gamma_b/2$.

By expanding the initial state with respect to these nine
eigenoperators, one can then compute the evolution of the system
under the initial conditions given in part B of Sec.~IV.

When the initial state is
$\rho(0)=\left|0,e\right>\left<0,e\right|$ one obtains the
following density operator at time $t$
\begin{eqnarray}\label{rhorabi}
 &&\rho(t)=\left(1-\frac{1}{2}\mathrm{e}^{-\frac{\gamma_a}{2}t}-\frac{1}{2}\mathrm{e}^{-\frac{\gamma_b}{2}t}\right)
 \left|E_0\right>\left<E_0\right|\nonumber\\
\nonumber\\
 &&+\frac{1}{2}\mathrm{e}^{-\frac{\gamma_a}{2}t}\left|E_{1,-}\right>\left<E_{1,-}\right|+
 \frac{1}{2}\mathrm{e}^{-\frac{\gamma_b}{2}t}\left|E_{1,+}\right>\left<E_{1,+}\right|\nonumber\\
\nonumber\\
 &&-\frac{1}{2}\mathrm{e}^{-\frac{\gamma_a+\gamma_b}{4}t}\left(\mathrm{e}^{2i\Omega t}\left|E_{1,-}\right>\left<E_{1,+}\right|+
\mbox{h.c.}\right),
\end{eqnarray}
which allows us to express the population of the state
$\left|0,g\right>$ as follows
\begin{equation}\label{0g0gRabinew}
P_{0,g}(t)= \left<0,g\right|\rho(t)\left|0,g\right>=1-\frac{1}{2}\mathrm{e}^{-\frac{\gamma_a}{2}t}-\frac{1}{2}\mathrm{e}^{-\frac{\gamma_b}{2}t}.\\
\end{equation}
The population  of the state $\left|1,g\right>$ is given by
\begin{eqnarray}\label{1g1gRabinew}
&&P_{1,g}(t)= \left<1,g\right|\rho(t)\left|1,g\right>=\frac{1}{4}\left[\mathrm{e}^{-\frac{\gamma_a}{2}t}+\mathrm{e}^{-\frac{\gamma_b}{2}t}\right.\nonumber \\
\nonumber\\
 &&\left.-\mathrm{e}^{\left(2i\Omega-\frac{\gamma_a+\gamma_b}{4}\right)t}-\mathrm{e}^{\left(-2i\Omega-\frac{\gamma_a+\gamma_b}{4}\right)t}\right].
\end{eqnarray}
Using the two equations above one can calculate
$P_g(t)=\left<g\right|\rho(t)\left|g\right>=\left<0,g\right|\rho(t)\left|0,g\right>+\left<1,g\right|\rho(t)\left|1,g\right>$
given in Eq. (\ref{ggRabinew}).

Equation~(\ref{0g0gRabinew}) and Eq.~(\ref{1g1gRabinew}) are to be
compared with the corresponding quantities computed using the
phenomenological master equation, given by Eq.~(\ref{oldME}), and
considering the same initial condition. Following the same lines
one derives Eq.~(\ref{0g0gRabiold}) for the population of the
ground state $\left|0,g\right>$ and
\begin{eqnarray}\label{1g1gRabiold}
 &&P_{1,g}^{\rm ph}(t) = \left<1,g\right|\rho(t)\left|1,g\right>=
 \frac{8\Omega^2}{16\Omega^2-\gamma^2}\mathrm{e}^{-\frac{\gamma}{2}t}\nonumber\\
\nonumber\\
 &&-\frac{16\Omega^2}{4\left(16\Omega^2-\gamma^2\right)}\mathrm{e}^{\frac{-\gamma+\sqrt{\gamma^2-16\Omega^2}}{2}t}\nonumber\\
\nonumber\\
 &&-\frac{16\Omega^2}{4\left(16\Omega^2-\gamma^2\right)}\mathrm{e}^{\frac{-\gamma-\sqrt{\gamma^2-16\Omega^2}}{2}t},
\end{eqnarray}
for the population of state $\left|1,g\right>$.

In the same way one can derive the time evolution predicted by the
microscopic master equation~(\ref{equation1excitation}) when the
initial condition is
$\rho(0)=\left|E_{1,+}\right>\left<E_{1,+}\right|$:
\begin{eqnarray}\label{rhoBell}
 \rho(t)&=\left(1-\mathrm{e}^{-\frac{\gamma_b}{2}t}\right)\left|E_0\right>\left<E_0\right|\nonumber\\
\nonumber\\
 &+\mathrm{e}^{-\frac{\gamma_b}{2}t}\left|E_{1,+}\right>\left<E_{1,+}\right|,
\end{eqnarray}
which gives the following expressions for the populations of
states $\left|0,g\right>$ and $\left|1,g\right>$
\begin{eqnarray}\label{popBell}
 &&P_{0,g}(t)=\left<0,g\right|\rho(t)\left|0,g\right>=1-\mathrm{e}^{-\frac{\gamma_b}{2}t},\nonumber\\
\nonumber\\
 &&P_{1,g}(t)=
 \left<1,g\right|\rho(t)\left|1,g\right>=\frac{1}{2}\mathrm{e}^{-\frac{\gamma_b}{2}t}.
\end{eqnarray}

Once more, we must compare these expressions with the
corresponding ones obtained by means of the phenomenological
master equation given by Eq.~(\ref{oldME}). Using this equation
one obtains
\begin{eqnarray}\label{0g0gBellold}
 &&P_{0,g}^{\rm ph}(t)=\left<0,g\right|\rho(t)\left|0,g\right>=1-\frac{16\Omega^2}{16\Omega^2-\gamma^2}\mathrm{e}^{-\frac{\gamma}{2}t}\nonumber\\
\nonumber\\
  &&+\frac{\gamma^2}{2\left(16\Omega^2-\gamma^2\right)}\mathrm{e}^{\frac{-\gamma+\sqrt{\gamma^2-16\Omega^2}}{2}t}\nonumber\\
\nonumber\\
  &&+\frac{\gamma^2}{2\left(16\Omega^2-\gamma^2\right)}\mathrm{e}^{\frac{-\gamma-\sqrt{\gamma^2-16\Omega^2}}{2}t},
\end{eqnarray}
for the population of $\left|0,g\right>$, and
\begin{eqnarray}\label{1g1gBellold}
 &&P_{1,g}^{\rm ph}(t)=\left<1,g\right|\rho(t)\left|1,g\right>=\frac{8\Omega^2}{16\Omega^2-\gamma^2}\mathrm{e}^{-\frac{\gamma}{2}t}\nonumber\\
\nonumber\\
 &&-\frac{\gamma^2-\gamma\sqrt{\gamma^2-16\Omega^2}}{4\left(16\Omega^2-\gamma^2\right)}\mathrm{e}^{\frac{-\gamma+\sqrt{\gamma^2-16\Omega^2}}{2}t}\nonumber\\
\nonumber\\
 &&-\frac{\gamma^2+\gamma\sqrt{\gamma^2-16\Omega^2}2}{4\left(16\Omega^2-\gamma^2\right)}\mathrm{e}^{\frac{-\gamma-\sqrt{\gamma^2-16\Omega^2}}{2}t},
\end{eqnarray}
for the population of $\left|1,g\right>$.

\end{document}